\documentclass[prd,preprint,nofootinbib]{revtex4}
\usepackage{amssymb}
\usepackage{graphicx}

\begin{document}

\title{Topological black holes dressed with a conformally coupled scalar
field and electric charge}
\author{Cristi\'{a}n Mart\'{\i}nez}
\email{martinez@cecs.cl}
\affiliation{Centro~de~Estudios~Cient\'{\i}ficos~(CECS),~Casilla~1469,~Valdivia,~Chile.}
\author{Juan Pablo Staforelli}
\email{jp@cecs.cl}
\affiliation{Departamento de F\'{\i}sica, Universidad de Concepci\'{o}n, Casilla 160-C,
Concepci\'{o}n, Chile.\\
Departamento de F\'{\i}sica, P. Universidad Cat\'{o}lica de Chile, Casilla
306, Santiago 22, Chile.\\
Centro de Estudios Cient\'{\i}ficos (CECS), Casilla 1469, Valdivia, Chile.}
\author{Ricardo Troncoso}
\email{ratron@cecs.cl}
\affiliation{Centro~de~Estudios~Cient\'{\i}ficos~(CECS),~Casilla~1469,~Valdivia,~Chile.}

\begin{abstract}
Electrically charged solutions for gravity with a conformally coupled scalar
field are found in four dimensions in the presence of a cosmological
constant. If a quartic self-interaction term for the scalar field is
considered, there is a solution describing an asymptotically locally AdS
charged black hole dressed with a scalar field that is regular on and
outside the event horizon, which is a surface of negative constant
curvature. This black hole can have negative mass, which is bounded from
below for the extremal case, and its causal structure shows that the
solution describes a \textquotedblleft black hole inside a black hole". The
thermodynamics of the non-extremal black hole is analyzed in the grand
canonical ensemble. The entropy does not follow the area, and there is
an effective Newton constant which depends on the value of the scalar field
at the horizon. If the base manifold is locally flat, the solution has no
electric charge, and the scalar field has a vanishing stress-energy tensor
so that it dresses a locally AdS spacetime with a nut at the origin. In the
case of vanishing selfinteraction, the solutions also dress locally AdS
spacetimes, and if the base manifold is of negative constant curvature a
massless electrically charged hairy black hole is obtained. The
thermodynamics of this black hole is also analyzed. It is found that the
bounds for the black holes parameters in the conformal frame obtained from
requiring the entropy to be positive are mapped into the ones that guarantee
cosmic censorship in the Einstein frame.
\end{abstract}

\maketitle

\preprint{{\tiny CECS-PHY-05/18} }

\section{Introduction}

In four-dimensional asymptotically flat spacetimes, only horizons with the
topology of a sphere are compatible with a well defined causal structure for
black holes \cite{Friedman:1993ty}. However, this topological censorship can
be circumvented introducing a negative cosmological constant and a number of
black holes with flat or hyperbolic horizons have been reported in four and
higher dimensions \cite%
{Lemos:1994xp,Cai-Zhang,Vanzo:1997gw,Brill:1997mf,Birmingham:1998nr,Cai-Soh}%
. Moreover, this class of black holes are found in gravity theories
containing higher powers of the curvature \cite%
{BHscan,Aros:2000ij,Cai-GB,Dehghani}. The black hole thermodynamics is
drastically affected by the topology of the event horizon. In fact, the
Hawking-Page phase transition for the Schwarzschild-AdS black hole \cite%
{Hawking:1982dh} does not occur for asymptotically locally anti-de Sitter
black holes whose horizons have vanishing or negative constant curvature,
and they are thermodynamically --locally-- stable since they have a positive
heat capacity \cite{Birmingham:1998nr}. The inclusion of electric charge
further modifies the thermodynamical properties of black holes and a more
complex phase structure arises, allowing for an analogous of the
Hawking-Page phase transition \cite%
{Brill:1997mf,Cai-Soh,Chamblin:1999tk,Cai:2004pz}. Further developments
concerning black holes with nontrivial topology can be found in Refs. \cite%
{topo}.

On the other hand, one could naively think that the simplest source of
matter for a black hole would corresponds to a real scalar field. However,
the so called no-hair conjecture \cite%
{Ruffini-Wheeler,Bekenstein:1971hc,Teitelboim-No-hair}, which originally
stated that a black hole should be characterized only in terms of its mass,
angular momentum and electric charge, indicates that one should rule out
this kind of hairy black hole solutions (for recent discussions see e.g., 
\cite{review}). Nonetheless, this conjecture can be circumvented in
different ways. For instance, exact black hole solutions with minimally and
conformally coupled self-interacting scalar fields have been found in three 
\cite{Martinez:1996gn,Henneaux:2002wm} and four dimensions \cite%
{Martinez:2002ru,Martinez:2004nb,MT} in the presence of a cosmological
constant $\Lambda $. In the case of vanishing $\Lambda $, a four-dimensional
black hole is also known \cite{BBMB}, but the scalar field diverges at the
horizon. Numerical hairy black hole solutions of this sort have also been
found in Refs. \cite%
{Torii:2001pg,Winstanley:2002jt,HM1,Hertog-MaedaSTAB,Winstanley-CQG,Winstanley+Radu}%
.

In this article, four-dimensional electrically charged solutions for the
Einstein-Maxwell action with a conformally coupled scalar field and a
cosmological constant are found. A quartic self-interaction term for the
scalar field, which does not break the conformal invariance of its field
equation is considered. Solutions with a nonvanishing selfinteraction
coupling constant are discussed in section \ref{alpha-non-zero}, where the
base manifold is assumed to be a surface of constant curvature. In the
spherically symmetric case, the solution reduces to the one found in Ref. 
\cite{Martinez:2002ru}, which describes a hairy black hole provided the
cosmological constant is non negative. If the base manifold is a surface of
negative constant curvature, the solution corresponds to an asymptotically
locally AdS charged black hole dressed with a scalar field that is regular
on and outside the event horizon. This black hole can have negative mass,
which is bounded from below for the extremal case, and its causal structure
shows that the solution describes a \textquotedblleft black hole inside a
black hole". The case of a vanishing electric charge, reduces to the
solution found in Ref. \cite{Martinez:2004nb}, for which some novel
interesting features have been found.\footnote{%
For this hairy black hole, there is a critical temperature below which a
black hole in vacuum undergoes a spontaneous dressing up with a nontrivial
scalar field. Moreover, in this case they only admit a finite number of
quasinormal modes \cite{QN-G}. The solution is also expected to be
stable against perturbations \cite{Winstanley-CQG}, \cite{Hertog-Hollands},
and it can be uplifted to a solution of eleven-dimensional 
supergravity  \cite{Papa11}.} The thermodynamics of the non-extremal black hole is analyzed in
the grand canonical ensemble, and it is shown that the entropy does not
follow the area law. If the base manifold is locally flat, the solution is 
electrically neutral, and the scalar field has a vanishing stress-energy
tensor so that it dresses a locally AdS spacetime with a nut at the origin.

Section \ref{alpha_nule} is devoted to analyze the case of vanishing
self-interaction coupling constant. Nontrivial solutions are found,  which
curiously have a vanishing energy momentum tensor. The most interesting case
is the one for negative cosmological constant with a base manifold of
negative constant curvature. In this case the solution corresponds to a
locally AdS spacetime, and describes a massless electrically charged hairy
black hole is obtained, and its thermodynamics is also analyzed. Finally, in
section \ref{Einstein-Frame}, the mapping of these solutions to the Einstein
frame, where the scalar field becomes minimally coupled, is discussed. It is
found that the bounds for the black holes parameters in the conformal frame
which are obtained from requiring the entropy to be positive are mapped into
the ones that guarantee cosmic censorship in the Einstein frame.

\section{Einstein-Maxwell theory with a conformally coupled self-interacting
scalar field}

\label{section2}

We consider the Einstein-Maxwell system in four dimensions with a
cosmological constant $\Lambda $ and a real conformally coupled
self-interacting scalar field, described by the action 
\begin{eqnarray}
I[g_{\mu \nu },\phi ,A_{\mu }] &=&\int d^{4}x\sqrt{-g}\left[ \frac{%
R-2\Lambda }{16\pi G}-\frac{1}{2}g^{\mu \nu }\partial _{\mu }\phi \partial
_{\nu }\phi -\frac{1}{12}R\phi ^{2}-\alpha \phi ^{4}\right]   \nonumber \\
&&-\frac{1}{16\pi }\int d^{4}x\sqrt{-g}F^{\mu \nu }F_{\mu \nu },
\label{action}
\end{eqnarray}%
where $G$ is the Newton's constant and $\alpha $ is an arbitrary coupling
constant. The corresponding field equations are\footnote{%
It can be shown that the inclusion of the electromagnetic field does not
spoil the duality symmetry described in Ref. \cite{Shapiro}.}%
\begin{eqnarray}  \label{maxwell}
G_{\mu \nu }+\Lambda g_{\mu \nu } &=&8\pi G(T_{\mu \nu }^{\phi }+T_{\mu \nu
}^{\mathrm{em}}),  \label{ee} \\
\square \phi  &=&\frac{1}{6}R\phi +4\alpha \phi ^{3},  \label{kg} \\
\partial _{\nu }(\sqrt{-g}F^{\mu \nu }) &=&0,  \label{Maxwell-eqn}
\end{eqnarray}%
where $\square \equiv g^{\mu \nu }\nabla _{\mu }\nabla _{\nu }$, and the
energy-momentum tensor is given by the sum of the scalar field contribution%
\begin{equation}
T_{\mu \nu }^{\phi }=\partial _{\mu }\phi \partial _{\nu }\phi -\frac{1}{2}%
g_{\mu \nu }g^{\alpha \beta }\partial _{\alpha }\phi \partial _{\beta }\phi +%
\frac{1}{6}[g_{\mu \nu }\square -\nabla _{\mu }\nabla _{\nu }+G_{\mu \nu
}]\phi ^{2}-g_{\mu \nu }\alpha \phi ^{4},  \label{Tuvfield}
\end{equation}%
and the one for the electromagnetic field%
\begin{equation}
T_{\mu \nu }^{\mathrm{em}}=\frac{1}{4\pi }\left( F_{\mu \alpha }F_{\nu \beta
}-\frac{1}{4}g_{\mu \nu }F_{\gamma \alpha }F_{\delta \beta }g^{\gamma \delta
}\right) g^{\alpha \beta }.  \label{Tuvem}
\end{equation}%
Note that Eq. (\ref{kg}) can be obtained from Eqs. (\ref{ee}) and (\ref%
{Maxwell-eqn}) using the conservation of the energy-momentum tensor, and since
the scalar field is conformally coupled, which means that the total
energy-momentum tensor is traceless,  the Ricci scalar curvature is
constant, $R=4\Lambda $.

\section{Solutions with a nonvanishing self-interaction coupling constant $(%
\protect\alpha \neq 0)$}
\label{alpha-non-zero}

The field equations (\ref{ee}), (\ref{kg}) and (\ref{Maxwell-eqn}) admit an
exact static solution whose metric is given by%
\begin{equation}
ds^{2}=-\Big[-\frac{\Lambda r^{2}}{3}+\gamma \Big(1+\frac{G\mu }{r}\Big)^{2}%
\Big]dt^{2}+\Big[-\frac{\Lambda r^{2}}{3}+\gamma \Big(1+\frac{G\mu }{r}\Big)%
^{2}\Big]^{-1}dr^{2}+r^{2}d\sigma ^{2},  \label{genericalphasolution}
\end{equation}%
with $-\infty <t<\infty $ and $r>0$. Here $d\sigma ^{2}$ stands for the line
element of the two-dimensional base manifold $\Sigma $ which is assumed to
be compact, without boundary, and of constant curvature $\gamma $ that can
be normalized to $\pm 1,0$. This means that the surface $\Sigma $ is locally
isometric to the sphere $S^{2}$, flat space $\mathbb{R}^{2}$, or to the
hyperbolic manifold $H^{2}$ for $\gamma =+1,0,-1$, respectively.

The scalar field reads 
\begin{equation}
\phi =\sqrt{\frac{-\Lambda }{6\alpha }}\frac{G\mu }{r+G\mu },  \label{scalar}
\end{equation}%
which is real provided $\alpha $ and $\Lambda $ have opposite signs, and the
electromagnetic potential is given by%
\begin{equation}
A=-\frac{q}{r}dt\;.  \label{electric}
\end{equation}%
The integration constants $q$ and $\mu $ are not independent since they must
satisfy%
\begin{equation}
q^{2}=\gamma G\mu ^{2}\left( 1+\frac{2\pi \Lambda G}{9\alpha }\right) \;.
\label{ratio}
\end{equation}%
As it it shown below, they are related to the electric charge $Q$ and the
mass $M$, respectively as%
\begin{equation}
M=-\gamma \frac{\sigma }{4\pi }\mu \quad \mbox{and}\quad Q=\frac{\sigma }{%
4\pi }q,  \label{BlackHoleMass}
\end{equation}%
where $\sigma $ is the area of the base manifold $\Sigma $.

For $\gamma =1$, the only sensible choice for the base manifold $\Sigma $ is
the sphere $S^{2}$, since the other smooth manifold available is the real
projective space $\mathbb{R}P^{2}$, which however, is non orientable. This
case has been discussed in Ref. \cite{Martinez:2002ru}, where it was shown
that the solution describes a hairy black hole provided the cosmological
constant is non negative. In the vanishing cosmological constant limit, the
self coupling constant must vanish also, and the solution of Ref. \cite{BBMB}
is recovered. The thermodynamics and the stability for the black hole with
positive cosmological constant have been discussed in Refs. \cite{ThermodSBH}
and \cite{Instability}, respectively. Further aspects of this class of
solutions have been studied in \cite{Further-aspects}. The remaining cases, $%
\gamma =-1,0$, are analyzed below.

\subsection{Electrically charged hairy black hole with hyperbolic horizon $(%
\protect\gamma =-1)$}

Here we consider the case when the base manifold is of negative constant
curvature $\gamma =-1$. This means that $\Sigma $ must be of the form $%
\Sigma =H^{2}/\Gamma $, where $\Gamma $ is a freely acting discrete subgroup
of $O(2,1)$. This manifold has genus $g\geq 2$, and its area is given by%
\[
\sigma =4\pi (g-1)\ . 
\]%
In this case, when the cosmological constant is negative ($\Lambda =-3l^{-2}$%
), the metric (\ref{genericalphasolution}) takes the form%
\begin{equation}
ds^{2}=-\Big[\frac{r^{2}}{l^{2}}-\Big(1+\frac{G\mu }{r}\Big)^{2}\Big]dt^{2}+%
\Big[\frac{r^{2}}{l^{2}}-\Big(1+\frac{G\mu }{r}\Big)^{2}\Big]%
^{-1}dr^{2}+r^{2}d\sigma ^{2},  \label{bhsolution}
\end{equation}%
and describes an asymptotically locally AdS black hole. There is a unique
curvature singularity at the origin ($r=0$). The electromagnetic field is
given by (\ref{electric}), and the scalar field

\begin{equation}
\phi =\sqrt{\frac{1}{2\alpha l^{2}}}\frac{G\mu }{r+G\mu },  \label{scalar2}
\end{equation}%
has a simple pole at $r=-G\mu $ only if $\mu $ is negative, and it is real
for $\alpha >0.$

The components of the total energy-momentum tensor, corresponding to
addition of (\ref{Tuvfield}) and (\ref{Tuvem}) are 
\begin{equation}
T_{\nu }^{\mu }=\frac{G\mu ^{2}}{8\pi r^{4}}\mathrm{diag}(-1,-1,1,1)\ ,
\label{components}
\end{equation}%
\ which satisfies the dominant and strong energy conditions. In fact, the
energy-momentum tensor is of type I in the classification of Ref. \cite%
{Hawking-Ellis}.

Equation (\ref{ratio}) fixes the charge-to-mass ratio of the black hole, as
a function of the fundamental constants appearing in the action: $G,\Lambda $
and $\alpha $, and determines an upper bound for the self-interaction
coupling constant%
\begin{equation}
0<\alpha \leq \frac{2\pi G}{3l^{2}}\ .  \label{Alpha-Bound}
\end{equation}%
This bound is saturated when the electric charge vanishes, and in this case
one recovers the neutral solution discussed in the appendix of Ref. \cite%
{Martinez:2004nb}. It is worth pointing out that, since the Ricci scalar is constant,
the scalar field equation (\ref{kg}) has a \textquotedblleft mexican hat"
shaped effective self interaction potential, whose mass term is given by $%
-2l^{-2}$, satisfying the Breitenlohner-Freedman bound \cite{BF}, which has
been shown to hold also for the topology considered here \cite{QN-AMTZ}.

The scalar field and the electromagnetic potential are given by (\ref{scalar}%
) and (\ref{electric}), respectively. Note that the scalar field and the
electric potential cannot be switched off keeping the mass fixed. Thus, for
fixed mass and electric charge, the theory described by the action (\ref%
{action}) admits two branches of black hole solutions with the same geometry
at the boundary: the hairy one presented here and the hyperbolic version of
the Reissner-Nordstr\"{o}m-AdS solution found in \cite{Brill:1997mf}. Both
branches intersect only for $\mu =q=0$, when the scalar field vanishes.

The causal structure of the electrically charges hairy black hole with
hyperbolic horizon radically depends on the value of the mass as it is
described below.

\subsubsection{Hairy black hole with non-negative mass}

Hairy black holes with positive mass, $\mu >0$, possess a single event
horizon located at 
\begin{equation}
r_{+}=\frac{l}{2}\left( 1+\sqrt{1+4G\mu /l}\right) \;,  \label{horizon}
\end{equation}%
satisfying $r_{+}>l$, and the scalar field is regular everywhere.

For the massless black hole, $\mu =0$, the scalar and electromagnetic field
vanish, and the metric acquires a simple form 
\begin{equation}
ds^{2}=-\left( \frac{r^{2}}{l^{2}}-1\right) dt^{2}+\left( \frac{r^{2}}{l^{2}}%
-1\right) ^{-1}dr^{2}+r^{2}d\sigma ^{2}\,,  \label{muzero}
\end{equation}%
which corresponds to a negative constant curvature spacetime. It has been
shown that this spacetime admits Killing spinors only if $\Sigma $ belong to
certain class of noncompact surfaces \cite{Aros:2002rk}, and its stability
has been analyzed in \cite{Gibbons:2002pq}.

For non-negative mass, the causal structure coincides with the one for
Schwarzschild-AdS black hole, but at each point of the Penrose diagram the
sphere is replaced by the base manifold $\Sigma $.

\subsubsection{Hairy black hole with negative mass: A black hole inside a
black hole}

It is well-known that the mass for black holes with hyperbolic horizons is
bounded from below by a negative value \cite{Vanzo:1997gw,Brill:1997mf,Mann:1997jb}%
. The hairy black hole presented here shares the same feature, since in
order to avoid the appearance of a naked singularity, the mass is bounded
according to%
\begin{equation}
G\mu \geq -\frac{l}{4}\ .  \label{Allowed-range}
\end{equation}%
For the generic case, i.e., when the mass is within the range $-l/4G<\mu <0$%
, the hairy black hole possesses three horizons located at%
\begin{eqnarray*}
r_{--} &=&\frac{l}{2}\left( -1+\sqrt{1-\frac{4G\mu }{l}}\right) \text{ }, \\
r_{-} &=&\frac{l}{2}\left( 1-\sqrt{1+\frac{4G\mu }{l}}\right) \text{ }, \\
r_{+} &=&\frac{l}{2}\left( 1+\sqrt{1+\frac{4G\mu }{l}}\right) \ ,
\end{eqnarray*}%
which satisfy $0<r_{--}<-G\mu <r_{-}<\frac{l}{2}<r_{+}$. The outermost and
innermost horizons, $r_{+}$ and $r_{--}$, respectively are event horizons,
and one can then say this metric describes ``a black hole inside a black
hole". The scalar field singularity, located at $-G\mu $, is timelike and
lies between the region bounded by $r_{--}$ and $r_{-}$, which is also
surrounded by $r_{+}$. However, note that the geometry is regular at this
timelike curve and hence, the stress-energy tensor is not singular there.
The causal structure can be constructed following the general prescriptions
indicated in \cite{Walker} and \cite{Klosh}. The resulting Penrose diagram
is shown in Fig. 1.

\begin{figure}
\label{p2}
\centering
\includegraphics[width=9cm]{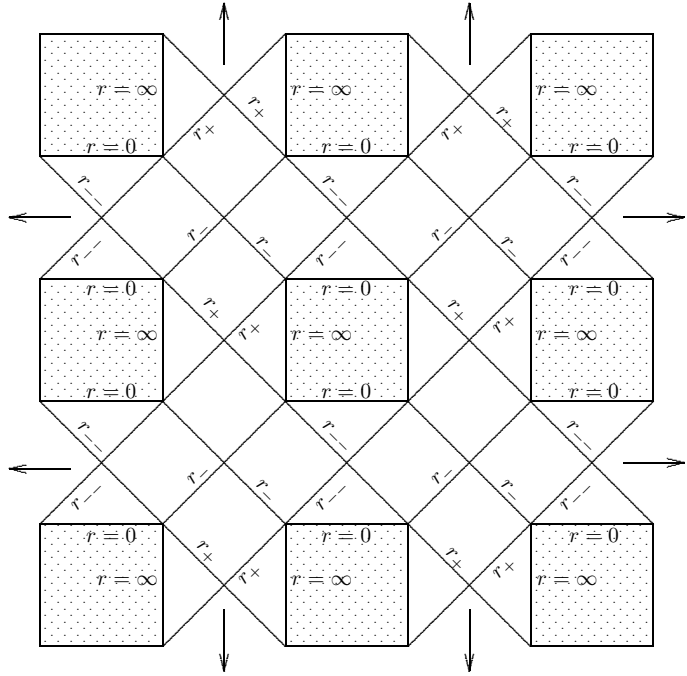}
\caption{Penrose diagram for the generic case:$ -l/4G<\mu<0 $. The dashed squared regions must be excised from the diagram.}
\end{figure}

The causal structure shows that the outermost region is causally similar to
the one for a Schwarzschild-AdS black hole. However, a timelike observer%
\footnote{%
We mean a test particle that follows a timelike curve.} who decide to cross
the outermost even horizon could eventually have a less disastrous fate.
Indeed at the first stage, the observer cannot avoid to go across $r_{-}$,
but nevertheless, once he reaches the region between $r_{-}$ and $r_{--}$,
he is able to ``switch on the rockets" and then eventually decide in what
region he will last his last days. This region is bounded by the innermost
event horizon $r_{--}$, and by $r_{-}$, which for him in practice is like a
cosmological horizon. Note that the timelike singularity of the scalar field
lies within this region, but as the geometry is regular there, test
observers cannot feel it\footnote{%
However, test fields that couple to the scalar field are expected to have an
ill-defined behavior there. Nonetheless, the scalar field singularity is
surrounded by ouertmost event horizon $r_{+}$.}. Therefore, the worst
possibility for the pointlike observer is to fall into the inner black hole
crossing $r_{--}$, inevitably finishing the trip in the future spacelike
singularity at $r=0$. Other possibility is to remain within the same region,
i.e., to avoid falling into the inner black hole without going across $r_{-}$%
. In this case, the trip finishes at the timelike future infinity located at
the corresponding bottom vertex of the excised squared region. The last
possibility is to irreversibly escape from this region by going across $%
r_{-} $. In this case, the observer cannot avoid to trespass back $r_{+}$
reaching to a mirror copy of the original outermost region. The future
infinity for a timelike observer in this region is located at the top of the
corresponding upper vertex of the excised squared region. In sum, a timelike
observer can go from one asymptotic region to a mirror copy of it without
facing any kind of geometric singularity along the trip.

It is also worth to point that a radial null geodesic is able to travel from
inside the inner white hole to $r=\infty $ in the outermost region.

The extremal case occurs for $G\mu =-l/4$, where the roots $r_{+}$ and $%
r_{-} $ coalesce taking the value $r_{+}=r_{-}=l/2>-G\mu >r_{--}>0$. Thus,
in this case, the solution describes ``a black hole inside an extremal black
hole". Its Penrose diagram is shown in Fig. 2.

\begin{figure}
\label{extreme}
\centering
\includegraphics[width=9cm]{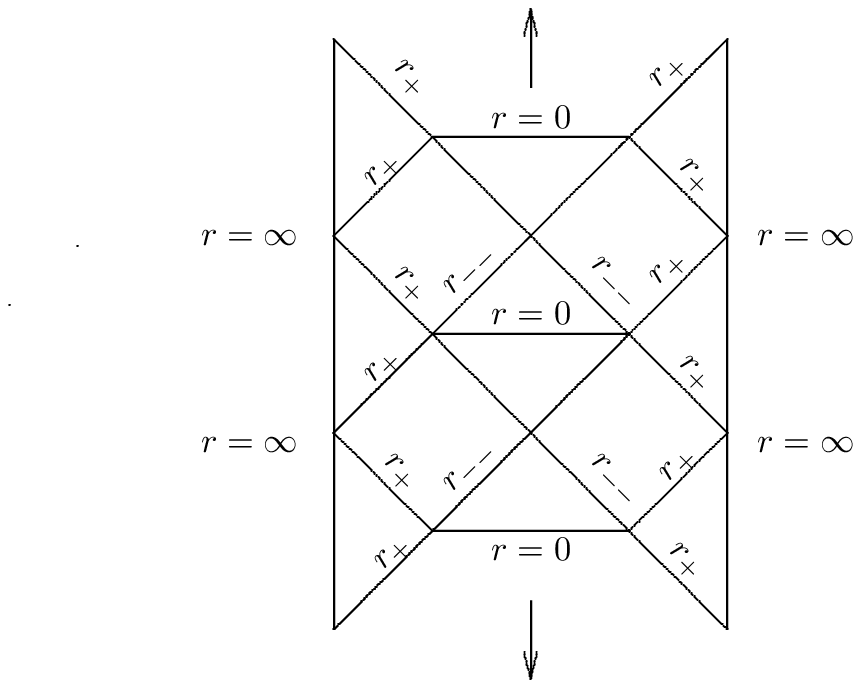}
\caption{Penrose diagram for $G\mu= -l/4$. This case corresponds to an extremal black hole.}
\end{figure}

From the causal structure, one can see that for the extreme case, a timelike
observer can also travel between two different asymptotic regions without
seeing singularities in the geometry along the path.

Finally, for the case $G\mu <-l/4$, there is a single event horizon located
at $r_{--}<-G\mu $, so that the geometric singularity at the origin is
protected, but the timelike singularity of scalar field is naked since is
accessible to observers in the asymptotic region. Note that, although this
singularity is not of geometrical origin, the expected ill behavior of
fields that couple to the scalar field would be measured at the asymptotic
region. In this sense, cosmic censorship is violated, and hence, this case
should not be regarded as a physical solution.

\subsubsection{Thermodynamics}

\label{thermo}

For non negative masses, the non-extremal black hole solution discussed in
the previous subsection possesses a single event horizon, and hence its
thermodynamics can be suitably analyzed following Euclidean methods taking
care of the nonminimal coupling of the scalar field with gravity. This
approach can also be applied for the allowed range of negative masses under
certain proper assumptions.

In the Euclidean method, the partition function for a thermodynamical
ensemble is identified with the Euclidean path integral in the saddle point
approximation around the Euclidean continuation of the classical solution 
\cite{Gibbons:1976ue}. The Euclidean continuation is obtained making $%
t\rightarrow it$, where the imaginary time has period $\beta $, which is
identified with the inverse temperature. Thus, for a thermodynamical
ensemble the action is related with free energy $\mathcal{F}$ as $I=\beta 
\mathcal{F}$.

In order to determine the suitable boundary terms that makes the Euclidean
action to be finite, it is enough to consider the minisuperspace of static
Euclidean metrics 
\begin{equation}
ds^{2}=N(r)^{2}f(r)^{2}dt^{2}+f(r)^{-2}dr^{2}+r^{2}d\sigma ^{2},  \label{dse}
\end{equation}%
with $0\leq t<\beta $, $r\geq r_{+}$. The scalar and electromagnetic fields
are assumed to be of the form $\phi =\phi (r)$, $A=A_{t}(r)dt$,
respectively. The reduced action principle is then%
\begin{equation}
I=\frac{\beta \,\sigma }{4\pi }\int_{r_{+}}^{\infty }(N(r)\mathcal{H}%
(r)+A_{t}p^{\prime })dr+B\ ,  \label{rhaction}
\end{equation}%
where $B$ is a surface term, $p:=\frac{r^{2}}{N}A_{t}^{\prime }\ $, the
symbol $^{\prime }$ denotes $d/dr$, and the reduced Hamiltonian $\mathcal{H}$
is given by 
\begin{eqnarray}
\mathcal{H} &=&\frac{r^{2}}{2G}\left[ 8\pi G\left\{ \frac{1}{6}f^{2}(\phi
^{\prime })^{2}-\frac{1}{6}\phi \phi ^{\prime }\left[ (f^{2})^{\prime }+%
\frac{4}{r}f^{2}\right] -\frac{1}{3}\phi f^{2}\phi ^{\prime \prime }+\alpha
\phi ^{4}\right\} \right.  \nonumber \\
&&+\left. \left( 1-\frac{4\pi G}{3}\phi ^{2}\right) \left( \frac{%
(f^{2})^{\prime }}{r}-\frac{1}{r^{2}}(-1-f^{2})\right) +\frac{Gp^{2}}{r^{4}}%
+\Lambda \right] \ .
\end{eqnarray}%
Since the Euclidean solution satisfies the constraint $\mathcal{H}=0$ and
the Gauss law $p^{\prime }=0$, the action evaluated on the classical
solution is just given by the boundary term $B$. In what follows, we work in
the grand canonical ensemble, so that we consider variations of the action
keeping fixed the temperature and the \textquotedblleft voltage", i.e., $%
\beta $ and $\Phi =A_{t}(\infty )-A_{t}(r_{+})$ are constants without
variation. The temperature is fixed requiring regularity of the metric at
the horizon, which yields%
\begin{equation}
\left. \beta (N(r)(f^{2}(r))^{\prime })\right\vert _{r=r_{+}}=4\pi \;.
\label{regular}
\end{equation}%
For the black hole (\ref{bhsolution}) one obtains that%
\begin{equation}
\beta ^{-1}=\frac{1}{2\pi l}\Big(\frac{2r_{+}}{l}-1\Big).
\end{equation}%
This boundary term is determined requiring the action (\ref{rhaction}) to
attain extremum, i.e., $\delta I=0$ within the class of fields considered
here \cite{Regge:1974zd}. This implies that the variation of the boundary
term is given by%
\begin{eqnarray}
\delta B &=&-\frac{\beta \sigma }{8\pi G}\left[ Nr\left( 1-\frac{4\pi G}{3}%
\phi ^{2}\right) \delta f^{2}+Nr^{2}\frac{4\pi G}{3}\left[ \phi \left(
(f^{2})^{\prime }\delta \phi -\phi ^{\prime }\delta f^{2}\right)
+f^{2}\left( 4\phi ^{\prime }\delta \phi -2\phi \delta \phi ^{\prime
}\right) \right] \right] _{r_{+}}^{\infty }  \nonumber \\
&&-\left. \frac{\beta \sigma }{4\pi }A_{t}\delta p\right\vert
_{r_{+}}^{\infty }\equiv \delta B(\infty )-\delta B(r_{+})\,.  \label{varB}
\end{eqnarray}%
By virtue of the field equations, one obtains that $p(r)=q$, and the
\textquotedblleft lapse" is chosen as $N(r)=1$.

The variation of the fields at the horizon acquire the form%
\begin{eqnarray}
\left. \delta p\right\vert _{r_{+}} &=&\delta q,  \label{varf} \\
\left. \delta f^{2}\right\vert _{r_{+}} &=&-\left. (f^{2})^{\prime
}\right\vert _{r_{+}}\delta r_{+}, \\
\left. \delta \phi \right\vert _{r_{+}} &=&\delta \phi (r_{+})-\left. \phi
^{\prime }\right\vert _{r_{+}}\delta r_{+}.  \label{varphih}
\end{eqnarray}%
It is useful to define the \textquotedblleft effective Newton constant" $%
\tilde{G}\ $as 
\begin{equation}
\frac{1}{\tilde{G}}=\frac{1}{G}\left( 1-\frac{4\pi G}{3}\phi ^{2}\right) \,.
\label{Effective-Newton}
\end{equation}%
Thus, replacing (\ref{varf})-(\ref{varphih}) in (\ref{varB}), we obtain that
the surface term at the horizon is%
\begin{equation}
\delta B(r_{+})=\delta \left( \frac{A_{+}}{4\tilde{G}_{+}}\right) -\beta 
\frac{\sigma }{4\pi }[A_{t}(r_{+})]\delta q\,,
\end{equation}%
where $A_{+}=\sigma r_{+}^{2}$ is the horizon area, and $\tilde{G}_{+}$ is
the effective Newton constant at the horizon. The boundary term can then be
integrated as%
\begin{equation}
B(r_{+})=\frac{A_{+}}{4\tilde{G}_{+}}+\frac{\beta \sigma }{4\pi }\Phi q\,.
\end{equation}%
The variation of fields at infinity are 
\begin{eqnarray}
\left. \delta p\right\vert _{\infty } &=&\delta q, \\
\left. \delta f^{2}\right\vert _{\infty } &=&-\frac{2G\delta \mu }{r}%
+O(r^{-2}), \\
\left. \delta \phi \right\vert _{\infty } &=&\sqrt{-\frac{\Lambda }{6\alpha }%
}\Big(\frac{1}{r}-\frac{2G\mu }{r^{2}}\Big)G\delta \mu +O(r^{-3}), \\
\left. \delta \phi ^{\prime }\right\vert _{\infty } &=&-\sqrt{-\frac{\Lambda 
}{6\alpha }}\Big(\frac{1}{r^{2}}-\frac{4G\mu }{r^{3}}\Big)G\delta \mu
+O(r^{-4})\ ,
\end{eqnarray}%
from which one is able to determine the remaining boundary term. Using Eq. (%
\ref{varB}), the variation of the surface term at infinity is found to be%
\begin{equation}
\delta B(\infty )=\frac{\beta \sigma }{4\pi }\delta \mu \ ,
\end{equation}%
which, since that $\beta $ is fixed, can be integrated as%
\begin{equation}
B(\infty )=\frac{\beta \sigma }{4\pi }\mu \ .
\end{equation}%
Finally, the finite Euclidean action reads 
\begin{equation}
I=B(\infty )-B(r_{+})=\frac{\beta \sigma }{4\pi }\mu -\frac{A_{+}}{4\tilde{G}%
_{+}}-\frac{\beta \sigma }{4\pi }\Phi q\ ,  \label{action_onshell}
\end{equation}%
up to an arbitrary additive constant.

Since the Euclidean action is related with the free energy in the grand
canonical ensemble, the mass $M$, the electric charge $Q$ and the entropy $S$
can be found using the familiar thermodynamical relations: 
\begin{eqnarray}
M &=&\Big(\frac{\partial }{\partial \beta }-\beta ^{-1}\Phi \frac{\partial }{%
\partial \Phi }\Big)I=\frac{\sigma }{4\pi }\mu , \\
Q &=&-\beta ^{-1}\frac{\partial I}{\partial \Phi }=\frac{\sigma }{4\pi }q, \\
S &=&\Big(\beta \frac{\partial }{\partial \beta }-1\Big)I=\frac{A_{+}}{4%
\tilde{G}_{+}}.
\end{eqnarray}

Note that due to the nonminimal coupling of the scalar field with spacetime
geometry, the entropy does not satisfy the area law, as in Refs. \cite%
{Martinez:1996gn,Henneaux:2002wm,area}.

For the neutral black hole ($Q=0$), since the effective Newton constant
satisfy $\tilde{G}_{+}>0$, the entropy is always positive. However, for the
electrically charged black hole, the positivity of the entropy implies that
the mass must be bounded according to%
\begin{equation}
-\frac{\sqrt{\frac{3l^{2}\alpha }{2\pi G}}}{(1+\sqrt{\frac{3l^{2}\alpha }{%
2\pi G}})^{2}}<\frac{\mu G}{l}<\frac{\sqrt{\frac{3l^{2}\alpha }{2\pi G}}}{(1-%
\sqrt{\frac{3l^{2}\alpha }{2\pi G}})^{2}}\text{\ }.  \label{Bound}
\end{equation}%
It is worth to point out that in general, if the mass is in the allowed
range (\ref{Allowed-range}), but does not satisfy the bound (\ref{Bound}),
the effective Newton constant at the horizon becomes negative, which
although the geometry is regular on and outside the event horizon, would
lead to an unphysical negative entropy. This bound, has a natural geometric
interpretation in the Einstein frame as explained in section \ref%
{Einstein-Frame}.

For the black holes with negative mass some comments are in order, since as
the solution was found to describe ``a black hole inside a black hole", we
are facing a situation where the object possesses two event horizons, and
therefore, extrapolation of the thermodynamical quantities found here must
be taken with care. Indeed, dealing with the thermodynamics of black holes
possessing more than one radiating horizon, as for the Schwarzschild-de
Sitter solution, is a subtle issue, and some results have been obtained
considering either of the two horizons as a boundary (for a recent treatment
see Ref. \cite{Gomberoff-Teitelboim}). For negative masses, as discussed
previously, the causal structure shows that radial null geodesics are able
to travel from inside the inner white hole to the outermost region, so that
in principle, observers in the asymptotic region would be able to receive an
additional contribution to the Hawking radiation coming from the inner black
hole if any. This issue is out of the scope of this work, and we only
consider the thermodynamics associated to the Hawking radiation coming from
the outermost event horizon. Under this assumption, the Euclidean approach
works properly once the inner region is removed.

\subsection{Hairy gravitationally stealth solution with locally flat base
manifold $(\protect\gamma =0)$}

Let us consider now the case when the base manifold is of vanishing
curvature, i.e., for $\gamma =0$. In this case, the electromagnetic field
vanishes, and when the cosmological constant is negative ($\Lambda =-3l^{-2}$%
), the metric (\ref{genericalphasolution}) acquires the form

\begin{equation}
ds^{2}=-\frac{r^{2}}{l^{2}}dt^{2}+\frac{l^{2}}{r^{2}}dr^{2}+r^{2}d\sigma
^{2}\ ,  \label{LAdSgamma0}
\end{equation}%
where the surface $\Sigma $ is locally isometric to flat space $\mathbb{R}%
^{2}$. The scalar field is given by Eq. (\ref{scalar2}), which is real
provided $\alpha $ is positive. In this case, the mass vanishes and hence
the integration constant $\mu $ is a \textquotedblleft pure hair" modulus
parameter. The scalar field is regular everywhere for positive $\mu $.

This solution is the four-dimensional generalization of the
\textquotedblleft nullnut" solution found in Ref. \cite{GMT} in the
conformal frame. Indeed, in this case the metric is locally AdS spacetime,
and since the timelike Killing vector vanishes at the origin, it has nut on
the null curve $r=0$. Its causal structure is then the same as the one for
the massless BTZ\ black hole \cite{BHTZ}.

As it occurs for its three-dimensional counterpart, the stress-energy tensor
vanishes. The effect of having nontrivial matter fields with a vanishing
total energy momentum tensor have also been discussed for flat spacetime in 
\cite{Cheshire,Robinson} and for three-dimensional gravity with negative
cosmological constant in Refs. \cite{GMT,Natsuume+Eloy+HMTZ2+1+Stealth}.
This effect has also been discussed for different setups in Refs. \cite%
{Mokhtar+Olivera+Demir}. In the next Section, new solutions of this sort are
found when the self-interaction coupling constant vanishes.

\section{Gravitationally stealth solutions with a vanishing self-interaction
coupling constant $(\protect\alpha =0)$}

\label{alpha_nule}

In the case of vanishing self-interaction coupling constant, i.e., for $%
\alpha =0$, we find a massless solution dressed with a non-trivial scalar
field. The metric reads

\begin{equation}
ds^{2}=-\left[ -\frac{\Lambda }{3}r^{2}+\gamma \right] dt^{2}+\left[ -\frac{%
\Lambda }{3}r^{2}+\gamma \right] ^{-1}dr^{2}+r^{2}d\sigma ^{2}\;,
\label{metric-alpha0}
\end{equation}%
the scalar field is given by 
\begin{equation}
\phi =\frac{A}{r}  \label{Scalar-alpha0}
\end{equation}%
and the electromagnetic field is%
\begin{equation}
F_{rt}=\partial _{r}A_{t}=\frac{q}{r^{2}}\,,  \label{electric-alpha0}
\end{equation}%
where the electric charge is related to the integration constant $A$ as

\[
q^{2}=-\frac{4\pi }{3}\gamma A^{2}\ . 
\]%
For this class of solutions, this last relation excludes the spherically
symmetric case ($\gamma =1$), and consequently, the metric is well behaved
only for negative cosmological constant $\Lambda =-3l^{-2}$. Thus, the
metric is a locally AdS spacetime, and therefore the solutions dress these
spaces since they have a vanishing energy-momentum tensor with non-trivial
matter fields.

When the curvature of the base manifold vanishes, i.e., for $\gamma =0$, the
metric coincides with the case discussed in the previous section (\ref%
{LAdSgamma0}), the electric charge vanishes, and the scalar field (\ref%
{Scalar-alpha0}) is singular on the null curve $r=0$.

In the case when the base manifold is of negative constant curvature $\gamma
=-1$, the solution describes an electrically charged hairy black hole. The
scalar and electric fields are given by (\ref{Scalar-alpha0}), and (\ref%
{electric-alpha0}), respectively, and the metric coincides with the one in
Eq. (\ref{muzero}), as in the case on nonvanishing self interaction coupling
constant. However, in this case the scalar field (\ref{Scalar-alpha0}) does
not vanish, and its singularity at the origin is hidden by the event horizon
located at $r_{+}=l$.

The thermodynamics of this object can be readily found from the results
obtained in section \ref{thermo}, and the mass, electric charge and the
entropy are found to be,

\begin{eqnarray*}
M &=&0\ , \\
Q &=&\frac{\sigma }{4\pi }q\ , \\
S &=&\frac{A_{+}}{4\tilde{G}_{+}}=\frac{\sigma l^{2}}{4}\left( \frac{1}{G}-%
\frac{q^{2}}{l^{2}}\right) \ ,
\end{eqnarray*}%
respectively.

Note that if the electric and scalar fields were absent, the entropy for
this black hole would be given by $S=\frac{\sigma l^{2}}{4G}$. This means
that the Euclidean action, and hence the entropy, are sensitive to the
presence of the gravitationally stealth hair.

The positivity of the entropy implies that the electric charge is bounded as%
\begin{equation}
\frac{Gq^{2}}{l^{2}}<1\text{\ }.  \label{Bound2}
\end{equation}%
If this bound is violated, the effective Newton constant at the horizon
becomes negative. Note that although the geometry is regular on and outside
the event horizon, this would yield an unphysical negative entropy. This
bound, was obtained in \cite{MT} in the Einstein frame where it has a
natural geometric interpretation.

\section{Mapping to the Einstein frame: Solutions with a minimally coupled
self-interacting scalar field}
\label{Einstein-Frame}

It is a well-known fact that there is a map between conformally and
minimally coupled scalar fields. For the action principle considered here (%
\ref{action}), the precise map is obtained as follows:

Performing a conformal transformation, followed of a scalar field
redefinition of the form 
\begin{equation}
\hat{g}_{\mu \nu }=\left( 1-\frac{4\pi G}{3}\phi ^{2}\right) g_{\mu \nu
}\;,\qquad \Psi =\sqrt{\frac{3}{4\pi G}}\tanh ^{-1}\sqrt{\frac{4\pi G}{3}}%
\phi \;,  \label{ChangingFrames}
\end{equation}%
the action (\ref{action}) is mapped to the Einstein frame 
\begin{equation}
I[\hat{g}_{\mu \nu },\Psi ,A_{\mu }]=\int d^{4}x\sqrt{-\hat{g}}\left[ \frac{%
\hat{R}}{16\pi G}-\frac{1}{2}\hat{g}^{\mu \nu }\partial _{\mu }\Psi \partial
_{\nu }\Psi -V(\Psi )\right] -\frac{1}{16\pi }\int d^{4}x\sqrt{-\hat{g}}%
F^{\mu \nu }F_{\mu \nu }\ ,  \label{Action-Einstein-frame}
\end{equation}%
where the self-interacting potential reads%
\[
V(\Psi )=\frac{\Lambda }{8\pi G}\left[ \cosh ^{4}\left( \sqrt{\frac{4\pi G}{3%
}}\Psi \right) +\frac{9\alpha }{2\pi G\Lambda }\sinh ^{4}\left( \sqrt{\frac{%
4\pi G}{3}}\Psi \right) \right] \ . 
\]%
Thus, any solution of the action in the conformal frame (\ref{action}) can
be mapped to a solution of the action in the Einstein frame (\ref%
{Action-Einstein-frame}) by means of the transformations (\ref%
{ChangingFrames}). This map was discussed in Ref. \cite{Martinez:2004nb}, in the absence
of the electromagnetic field.

Aplying (\ref{ChangingFrames}) to the black hole solution given by Eqs. (\ref%
{bhsolution}), (\ref{scalar2}), and (\ref{electric}) leads to an
electrically charged hairy black hole for gravity with a minimally coupled
self interacting scalar field, which for the case of vanishing electric
charge reduces to the one found in \cite{Martinez:2004nb}. It can also be shown that the
electrically charged black hole solution with a vanishing self interaction
given by Eqs. (\ref{metric-alpha0}) with $\Lambda =-3l^{-2}$, (\ref%
{Scalar-alpha0}), and (\ref{electric-alpha0}) can mapped to the one in Ref. 
\cite{MT} using (\ref{ChangingFrames}) and a coordinate transformation.

Note that this map is invertible provided the conformal factor in Eq. (\ref%
{ChangingFrames}) --which goes like the inverse of the effective Newton
constant (\ref{Effective-Newton})-- does not vanish. This means that the
hypersurfaces in the conformal frame where the conformal factor vanishes are
mapped into singularities of the geometry in the Einstein frame. Hence, in
order to avoid dynamical instabilities, the region that one maps from the
conformal to the Einstein frame is the one with positive effective Newton
constant. Consequently, only a subset of well-behaved black hole solutions
in the conformal frame can be mapped to well-behaved black holes in the
Einstein frame. This subset is then defined by the black holes in the
conformal frame for which the effective Newton constant becomes negative
only inside the event horizon; otherwise the black hole would be mapped to a
naked singularity in the Einstein frame. Therefore one concludes that the
bounds for the black holes parameters in the conformal frame obtained from
requiring the entropy to be positive, as in Eqs. (\ref{Bound}) and (\ref%
{Bound2}), coincide with the ones that guarantee cosmic censorship in
Einstein frame. This is due to the fact that the black hole entropy in the
conformal frame is shown to be a quarter of the horizon area divided by the
effective Newton constant at the horizon, which means that the class of
causally well-behaved black hole solutions in the conformal frame which are
regarded as unphysical because they would lead to a negative entropy, in the
Einstein frame are mapped to spacetimes with naked singularities, which are
discarded by cosmic censorship. It is worth pointing out that a dynamical
argument in the conformal frame is mapped into a geometrical one in the
Einstein frame.

As a final remark, it is worth pointing out that once the asymptotically
locally AdS solutions presented here are mapped to the Einstein frame, they
develop an asymptotic fall-off for the fields which is slower than the
standard one, as in Ref. \cite{Henneaux-Teitelboim}, for a localized
distribution of matter. As discussed in \cite{Henneaux:2002wm}, and further
developed in \cite{HMTZ-Log,Hertog-Maeda,HMTZ-D}, the presence of scalar
fields with a slow fall off gives rise to a strong back reaction that
relaxes the standard asymptotic form of the geometry, and it generates
additional contributions to the charges that depend explicitly on the scalar
fields at infinity which are not already present in the gravitational part.
These effects has also been discussed recently following different
approaches in \cite{Marolf,Glenn,Chen-Lu-Pope}.

\acknowledgments

We thank Jorge Zanelli, who participated in the initial stages of this work,
and Eloy Ay\'{o}n-Beato, I. Shapiro and Steven Willison for useful comments.
This research is partially funded by FONDECYT grants N%
${{}^o}$
1051064, 1051056, 1040921, 1061291. The generous support to Centro de
Estudios Cient\'{\i}ficos (CECS) by Empresas CMPC is also acknowledged. CECS
is a Millennium Science Institute and is  funded in part by grants from the
Millennium Science Initiative, Fundaci\'{o}n Andes and the Tinker Foundation.

\end{document}